\documentclass[sigconf]{acmart}

\AtBeginDocument{%
  \providecommand\BibTeX{{%
    \normalfont B\kern-0.5em{\scshape i\kern-0.25em b}\kern-0.8em\TeX}}}

\copyrightyear{2026}
\acmYear{2026}
\setcopyright{cc}
\setcctype{by}
\acmConference[CHI '26]{Proceedings of the 2026 CHI Conference on Human Factors in Computing Systems}{April 13--17, 2026}{Barcelona, Spain}
\acmBooktitle{Proceedings of the 2026 CHI Conference on Human Factors in Computing Systems (CHI '26), April 13--17, 2026, Barcelona, Spain}
\acmDOI{10.1145/3772318.3790297}
\acmISBN{979-8-4007-2278-3/2026/04}

\newenvironment{myquote}%
  {\list{}{\leftmargin=0.1in\rightmargin=0.1in}\item[]}%
  {\endlist}

\begin{document}

\title{The Promises and Perils of using LLMs for Effective Public Services}


\author{Erina Seh-Young Moon}
\affiliation{%
\institution{University of Toronto}
  \city{Toronto}
  \country{Canada}}
\email{erina.moon@mail.utoronto.ca}
\orcid{0000-0003-3233-9773}

\author{Matthew Tamura}
\affiliation{
\institution{University of Toronto}
  \city{Toronto}
  \country{Canada}
 }
\email{matthew.k.tamura@gmail.com}
\orcid{0009-0001-8674-3942}

\author{Angelina Zhai}
\affiliation{
\institution{Georgia Institute of Technology}
  \city{Atlanta}
  \country{United States}
 }
\email{angelina.zhai@gatech.edu}
\orcid{0009-0006-6628-4888}

\author{Nuzaira Habib}
\affiliation{
\institution{University of Toronto}
  \city{Toronto}
  \country{Canada}
 }
\email{nuzaira.habib@mail.utoronto.ca}
\orcid{0009-0007-7708-3683}

\author{Behnaz Shirazi}
\affiliation{
\institution{Child Welfare Institute, CAS of Toronto}
  \city{Toronto}
  \country{Canada}
 }
\email{bshirazi@torontocas.ca}
\orcid{0009-0005-5732-6129}

\author{Altaf Kassam}
\affiliation{
\institution{Child Welfare Institute, CAS of Toronto}
  \city{Toronto}
  \country{Canada}
 }
\email{akassam@torontocas.ca}
\orcid{0009-0000-2080-9382}

\author{Devansh Saxena}
\affiliation{%
\institution{University of Wisconsin-Madison}
  \city{Madison}
  \country{United States}}
\email{devansh.saxena@wisc.edu}
\orcid{0000-0001-5566-7409}

\author{Shion Guha}
\affiliation{%
\institution{University of Toronto}
  \city{Toronto}
  \country{Canada}
 }
\email{shion.guha@utoronto.ca}
\orcid{0000-0003-0073-2378}

\renewcommand{\shortauthors}{Moon et al.}

\begin{abstract}

Governments are the primary providers of essential public services and are responsible for delivering them effectively. In high-stakes decision-making domains such as child welfare (CW), agencies must protect children without unnecessarily prolonging a family’s engagement with the system. With growing optimism around AI, governments are pushing for its integration but concerns regarding feasibility and harms remain. Through collaborations with a large Canadian CW agency, we examined how LocalLLM and BERTopic models can track CW case progress. We demonstrate how the tools can potentially assist workers in opportunistically addressing gaps in their work by signaling case progress/deviations. And yet, we also show how they fail to detect case trajectories that require discretionary judgments grounded in social work training, areas where practitioners would actually want support to pre-emptively address substantive case concerns. We also provide a roadmap of future participatory directions to co-design language tools for/with the public sector.
\end{abstract}




\maketitle

\section{Introduction}

With growing optimism around the potential of AI to improve public operations and services, governments in North America are aggressively experimenting with and adopting AI tools within public agencies \cite{brookings, cohere, thestar_publicregistry, hjaltalin2024}. For example, the US Department of Homeland Security recently released a guide on how government officials can improve public service delivery through the responsible use of generative AI tools \cite{dhs_genai}. Additionally, Canada was the first country to implement a national AI strategy \cite{blair2024} and has recently partnered with a private company that develops LLMs to identify opportunities for AI to improve public sector services \cite{cbc_cohere}. Underlying this movement lies an assumption that AI tools such as LLMs can deliver significant efficiency and tangible productivity gains for public agencies while reducing wasteful governmental spending \cite{dodge, whitehouse, cbc_cohere}. However, what remains insufficiently explored is how specific applications of AI can improve current public sector practices. 

The HCI community has a long history of researching the public sector using computational text analysis tools and following the emergence of LLMs, interest in this area has grown significantly \cite{pang25}. Previously, Saxena et al. \cite{chi23paper} applied LDA topic modeling to deconstruct how risk is conceptualized within child welfare narratives, and Field et al. \cite{field23} examined whether natural language processing tools can exacerbate racial biases within these systems. Recently, Nelson et al. \cite{nelson2024designing} also explored the feasibility of using LLMs to summarize casenotes to assist homelessness caseworkers in their jobs. Over the years, SIGCHI research has moved from applying computational text analysis tools on narrative documents to understand street-level realities of the public sector to exploring how LLMs can be used on these texts to support workers \cite{gondimalla24, nelson2024designing, moongi24}.

Motivated by the growing interest of North American governments in adopting AI tools for the public sector \cite{brookings, cohere, thestar_publicregistry}, our study builds on prior research that has examined the potential of applying computational text analysis tools on social work narrative texts \cite{gondimalla24, nelson2024designing, saxena22, chi23paper, moongi24}. In this study, through collaborations with a large Canadian child welfare agency, HCI scholars and child welfare practitioners who work at the agency collaborated together to explore how AI tools can be used to address a key operational challenge for the agency, ensuring that clients do not remain engaged with the agency for longer than necessary. The agency was concerned that caseworkers were engaging with some families for unnecessarily long periods and wanted to explore how computational text analysis tools could be applied to casenote narratives to improve their operational practices. To this end, we sought to track a family’s progress in meeting predefined Service Plan goals in Regular casenotes using BERTopic \cite{grootendorst2022bertopic} and LocalLLMs (specifically, Llama 3.1 \cite{llama31}). Due to the highly sensitive nature of the data, we employed computational text analysis tools that can be run locally \cite{perron25}. In this study, we asked the following research questions:

\begin {itemize} 
  \item \textbf{RQ1:} How can language models be used to understand how case management goals are defined and tracked in social work?
  \item \textbf{RQ2:} How can LLMs assist caseworkers to track case progress goals?
  \item \textbf{RQ3:} How can HCI researchers engage with the public sector to design LLM tools that support public sector work processes?
\end{itemize}

This paper makes the following unique research contributions. 

\begin{itemize}

\item We showcase an example of how HCI researchers and child welfare professionals can come together to investigate how language models can be used to address an important operational challenge for a large child welfare agency. 

\item Drawing on two types of child welfare documents (Service Plans and Regular casenotes), we illustrate how caseworkers can use AI tools based on language models to track how child welfare case management goals are defined and are being met. 

\item We outline the opportunities and limitations of using LLMs within child welfare systems. We reveal how child welfare practitioners can take advantage of the scalable capabilities of LLMs to support the delivery of effective services to clients. However, we also highlight how social work is necessarily inherently interpretive and context-dependent, which means that LLMs cannot and should not replace street-level decision-making \cite{lipsky2010street}. 

\item We outline a roadmap for how HCI scholars can collaborate with the public sector to design AI tools for and with the public sector.

\end{itemize}

\begin{figure*}[]
\centering 
\includegraphics[scale=0.43]{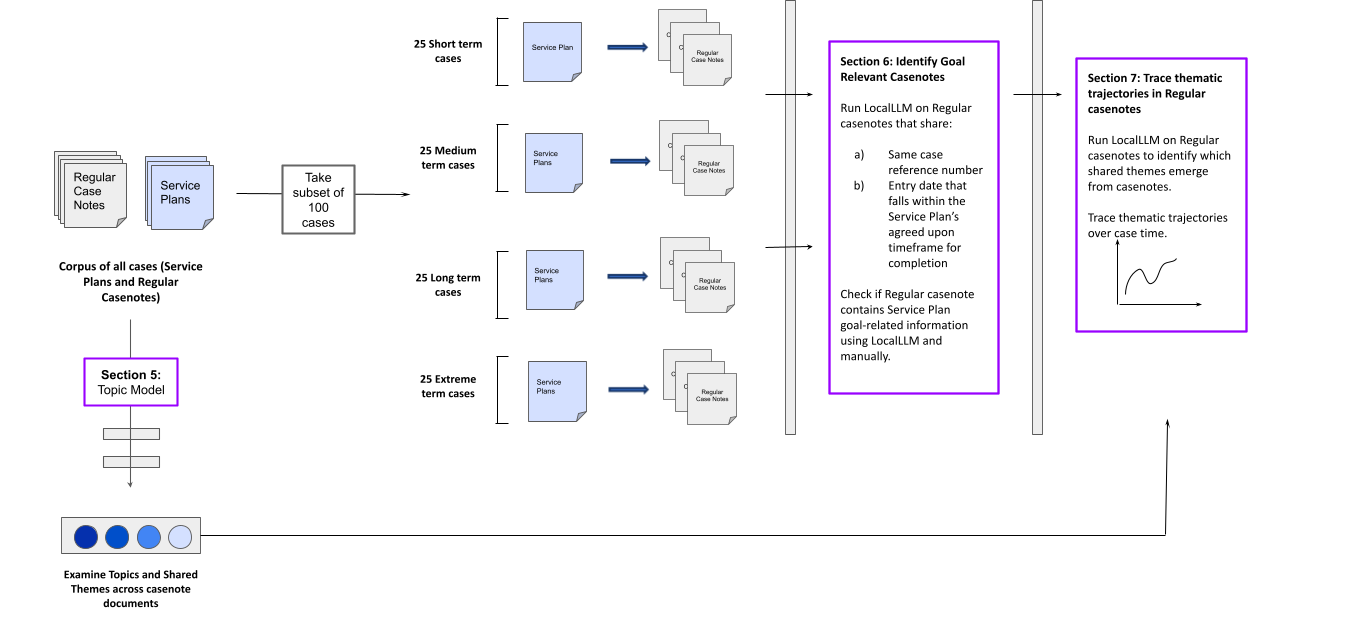}
\caption{Visualization of this study's analysis steps}
\label{fig:flowchart}
\end{figure*}

To address our RQs, we structured our study as a three-stage process, which we visualize in Figure \ref{fig:flowchart}. To address RQ1 (Section \ref{sec:study1}), we applied BERTopic on Service Plan goals and Regular casenote narratives. We extracted topics and elicited shared themes between the two document types to understand how the types of child welfare case management goals mentioned in the Service Plan are tracked and worked on throughout a case. We addressed RQ2 in two steps (Sections \ref{sec:study2} and \ref{sec:exp4}). We took a subset of our casenote dataset to apply a LocalLLM to identify which Regular casenotes contain information relevant to Service Plan goals. We manually labeled the casenotes as well to evaluate the LocalLLM's performance. Then, on the same subsetted casenote dataset, we used a LocalLLM to inquire which themes we had identified in the first step of the study (Section \ref{sec:study1}) appeared in Regular casenotes that had been labelled as containing Service Plan-relevant/irrelevant goals. Through these two stages, we could identify case progress-relevant documents and examine their narrative content. We then reflected on RQ3 in the Discussion section. 

\section{Reflection on Research Ethics}
We received approval from the Research Ethics Board of our university to use child welfare narrative documents for this research. We are deeply aware that beyond the REB guidelines, research using child welfare documents is morally complex and entangled with issues of oppression and surveillance. Considering that Canadian governments are pushing for greater AI adoption within the public sector, we hope our study can surface opportunities for and limitations on using AI within the child welfare space. When conducting this study, we considered the principles of public interest and closely collaborated with our co-authors at the child welfare agency to ensure that we were conducting research in the interest of the agency’s clients. We anonymized all personal information from the data and do not make any raw data public.

\section{Related Work}
\subsection{SIGCHI research on public sector data-driven tools}
SIGCHI researchers have a long history in critically researching public sector sociotechnical systems, including examining issues around algorithmic governance and public data \cite{kawakami2022, jin_reasonabledoubt24, moon25_datafication, kelly24}, introducing novel participatory HCI methodologies \cite{kuo23, gordon_jurylearning22, halperin23}, and designing human-centered computational tools that promote human agency and empowerment \cite{lee_webuildai19, gondimalla24, saxena_partic20, chi25_workshop}. Most relevant to this study, in recent years, the SIGCHI community has turned its attention to researching algorithmic decision-making tools that are built using the vast quantities of personal data collected by public agencies \cite{Levy_2021, chouldechova18, brown19, showkat_25}. Systematic literature reviews on the technical components of these tools have found that these decision-making tools are largely configured to solve prediction problems \cite{moon24, saxena2020human, kelly23, johnson22, showkat23}. Learning patterns in historical data using statistical and machine learning techniques, the majority of these tools are intended to classify people according to some value or risk they present to society so that those deemed ‘deserving' can be prioritized for public service delivery \cite{liu2025bridging, johnson22}. The overarching impetus for using these tools is to support or supplant human decision-making such that public services are delivered to clients in a more consistent, objective, and evidence-based manner \cite{liu2025bridging,  richardson21}. However, recent work by SIGCHI scholars has raised concerns on the validity and limited utility of these tools when deployed on the ground. For example, in studying predictive algorithms designed for adults and homelessness social service provision, Showkat et al. \cite{showkat23} and Reinmund et al. \cite{Reinmund2024} found that cost-saving and efficiency are central values promoted in these algorithms, which can result in the de-prioritization of important human values and an abstraction away from critical contextual client information. Furthermore, Moon and Guha \cite{moon24} and Saxena et al. \cite{saxena2021framework2} found that such predictive algorithms often fail to account for their resource-constrained deployment context. Accurate algorithmic prediction outputs do not necessarily translate to the successful delivery of public services because public resources, such as good foster homes or housing opportunities, are often very limited \cite{saxena2023algorithmic, eubanks2018automating, Reinmund2024, moon24, liu2024}. 

Recognizing the implementation gap between decisions rendered by predictive decision-making algorithms and the delivery of public services, researchers have called for the need to shift from a prediction-focused paradigm towards an intervention-focused paradigm – I.e., where public sector data should be used to effectively support public sector workers’ decision-making processes rather than designing predictive decision-making algorithms that solely focus on the efficient delivery of services \cite{moon24, gondimalla24, kawakami2022, saxena2021framework2, liu2025bridging, fox17socialjusticedesign}. Social service literature has long found that decision-making and the quality of public services delivered can be improved when organizations focus on how well an organization meets its predefined objectives \cite{LeRoux_Wright_2010, Ylonen_2023, Smith_1988, Clarkson_2008}. In line with this social work scholarship, SIGCHI researchers have proposed applying novel co-design methodologies such as comic boarding and voting to elicit stakeholder needs and preferences to design effective technical tools that center stakeholder (and in particular) client needs \cite{kuo23, gondimalla24, lee_webuildai19, gordon_jurylearning22, tang_failurecards24}. Furthermore, HCI scholars have also begun exploring the viability of applying computational text analysis techniques on narrative casenotes written by caseworkers to support public sector workers \cite{saxena22, chi23paper, gondimalla24, nelson2024designing}. Nelson et al. \cite{nelson2024designing} recently found that caseworkers found value in computational tools that can summarize casenotes to provide customized services for clients and Saxena et al. \cite{chi23paper} unveiled that casenotes can reveal critical and contextual case information, highlighting their potential as a valuable data source for developing decision-support tools for public sector workers.

\subsection{SIGCHI research on child welfare}

Recent SIGCHI research on child welfare systems has centered on understanding how technologies can empower and uplift the values of different groups engaged in CW services. Work in this domain has been wide-ranging. Some studies have focused on how CWS members engage with- and mediate- online technology. For example, Ammari et al. \cite{ammari25} examined how individuals in the foster care system find community on online Reddit communities. Furthermore, Caddle et al. \cite{ caddle22}, Badillo-Urquiola et al. \cite{karla19}, and Badillo-Urquiola et al. \cite{karla_24} have inquired how collaborative sociotechnical systems that engage foster parents and caseworkers can mitigate adolescent online safety risks. In recent years, a plethora of research has critically examined CWS algorithmic decision-making tools, which have grown in prominence across North American child welfare agencies \cite{saxena2020human, samant21}. Through a review of these tools \cite{saxena2020human} and community design workshops with stakeholders \cite{stapleton22, brown19}, the SIGCHI community has consistently found that CW algorithms narrowly focus on case deficits by using the target outcome – child maltreatment risk – rather than focusing on family strengths. The researchers \cite{saxena2020human, stapleton22, brown19} have argued this is problematic as the algorithmic targets diverge from the overarching goal to improve child welfare and instead lead to systemically biased and punitive outcomes for families. SIGCHI researchers have also conducted audits examining the technical underpinnings of these algorithms, which support the above-mentioned literature. For example, in reviewing different CW decision-making algorithms that predict child maltreatment risk, Gerchick et al. \cite{gerchick_23} and Moreau et al. \cite{moreau24} found CW models used protected attributes and racially biased features that reinforce systemic discrimination and stigmatization. Additionally, Saxena et al. \cite{saxena2022chilbw} found that the tools would use a family's cooperativeness with caseworkers to predict child maltreatment risk rather than assess CW interventions' effectiveness. On-the-ground studies examining how workers engage with these tools found that caseworkers are, in fact, aware of the tools' limitations and accordingly detect/mediate erroneous algorithmic outputs by gaming inputs going into the tools \cite{saxena2021framework2} or overriding erroneous recommendations \cite{kawakami2022, feng22, deart_20}. With increasing awareness of the limitations of current algorithms in improving child welfare outcomes, SIGCHI researchers have increasingly emphasized the need to move away from predictive, deficit-focused models \cite{stapleton22, moongi24, chi23paper} and instead towards designing fair, human-centered, holistic technical tools that support child welfare workers in making informed, strength-based decisions for families  \cite{kawakami2022, cheng21, saxena_partic20, saxena22}. 

\subsection{SIGCHI Computational text analysis research on the public sector}
With the growing affordances from computational text analysis techniques, the SIGCHI community has increasingly explored how language tools can be used to study sociotechnical systems and elicit contextual information  \cite{gondimalla24, ammari25, abebe_mortality, chancellor16}. For example, Abebe et al. \cite{abebe_mortality} and Chancellor et al. \cite{chancellor16} found that applying topic models on social media data can provide predictive information on maternal mortality rates and mental illness severity. Relatedly, Antoniak et al. \cite{antoniak19} found running topic models on birth stories on Reddit can reveal aggregated patterns of typical narrative sequences and also rich and unique information on individual birth experiences and power relationships between personas \cite{antoniak19}. A growing body of scholarship has also explored the viability and opportunities of applying computational text analysis on public sector narrative casenote data \cite{saxena22, chi23paper, moongi24, field23, nelson2024designing}. In many public sector domains, caseworkers document copious unstructured narratives that record their observations, relevant details, and interactions with relevant parties \cite{Tracey_record_2023, saxena22, nielsen2023}. By applying topic models on narrative casenotes, researchers have shown the viability of applying computational text analysis on these documents to elicit patterns of invisible work conducted by caseworkers that are unaccounted for in ethnographic studies \cite{saxena22} and opportunities to infer how procedural and transitory factors impact bureaucratic street-level decisions \cite{chi23paper}. Furthermore, Field et al. \cite{field23} have explored the possibility of examining racial disparities in casenotes using word statistics and named entity recognition analysis, and Moon et al. \cite{moongi24} have identified limitations in using casenotes to predict future outcomes through predictive validity assessments.

Most recently, there has been growing interest within and beyond SIGCHI in exploring how LLMs can be integrated within the public sector to support public sector workers \cite{pang25, nelson2024designing, perron25, perron25_wordembeddings, perron_rag2025, baez_2025, Singer_2023}. For example, within HCI literature, through design sessions with homelessness caseworkers, Gondimalla et al. \cite{gondimalla24} and Nelson et al. \cite{nelson2024designing} found that workers are interested in the development of tools, including LLMs, that allow them to effectively use information from casenotes to support informed decision-making for clients. While public agencies collect vast quantities of narrative data on clients on electronic information systems, social work scholarship has consistently found that caseworkers struggle to parse through dense casenote documents to identify core issues impacting clients and quickly access relevant information \cite{Salovaara_Ylonen_2022, Huuskonen_Vakkari_2015, Pithouse_2012}. LLMs appear to offer new technical opportunities to distill and surface key information. Beyond SIGCHI, Perron et al. \cite{perron25} and Stoll et al. \cite{stoll2025} evaluated the ability of LocalLLMs in extracting child welfare case factors from unstructured child welfare narratives, and Báez et al. \cite{ baez_2025} and Singer et al. \cite{Singer_2023} examined how social work educators can leverage LLMs to enhance social work education. 

Our literature review shows there has been growing interest in SIGCHI and beyond to utilize computational text analysis techniques on public sector data - particularly casenote data - to support caseworkers in their work. However, there remain immense concerns around using LLMs on highly-sensitive data, the subjectivity of their outputs, and the risk of human over-reliance on them in high-stakes contexts \cite{perron25, paula25, bo25}. As governments aggressively explore how public agencies can integrate LLMs within the public sector, we applied BERTopic and local language models on child welfare casenotes to examine how computational text analysis tools can and cannot aid caseworkers in effectively delivering child welfare services for families.

\section{Research Context}

In this study, we partnered with a non-profit child welfare agency that serves a large metropolitan area in Canada. The organization is regulated and governed by the provincial child welfare ministry with the mandate to protect children and youth from abuse and neglect by providing risk assessment, family guidance and counselling, and permanency planning services. The organization is always interested in improving its operational decision-making processes to enable caseworkers to deliver timely, targeted services while minimizing unnecessary family involvement with the system. While acknowledging that families may require CW support to mitigate the risk of future harm to children, the organization also noted that prolonged involvement with the system can impose an intrusive and stressful burden on families and divert caseworkers from providing support to other families in need \cite{merritt, copeland2021s}. Therefore, through examinations of child welfare casenotes, the organization wanted to gain an in-depth understanding of how child welfare case goals were defined, tracked, and achieved, as well as why some cases exceeded average timeframes for case resolution. Following provincial child protection guidelines, caseworkers at the agency document all client-related information, including internal and external discussions, services delivered, and observations about the family, in casenotes. As a result, child welfare casenotes provide a rich source of contextual insights into a family’s experience within the child welfare system \cite{saxena22, chi23paper, moongi24}. For this study, we focused on two types of narrative documents documented by caseworkers, Service Plans and Regular Casenotes for families receiving Ongoing Case Management Services. We provide further detail on this in the following paragraphs.

\begin{figure*}[]
\centering 
\includegraphics[scale=0.3]{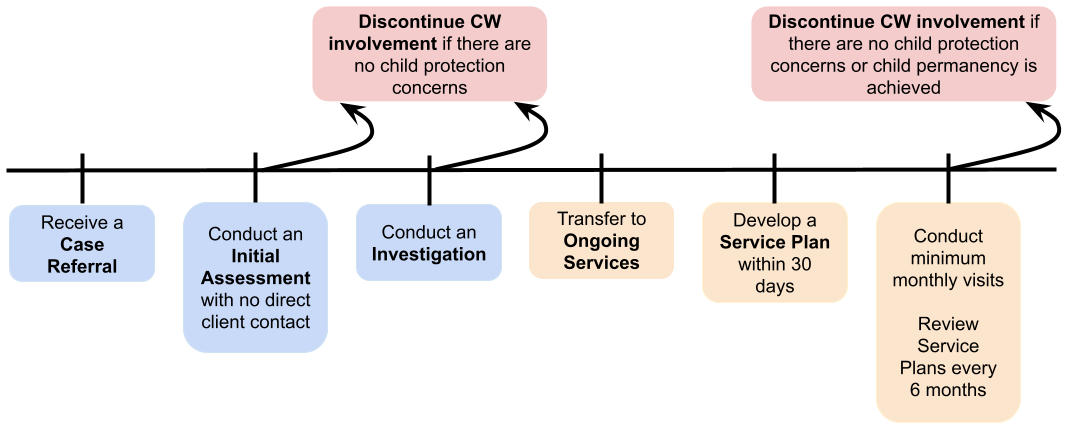}
\caption{Simplified case flow diagram within the agency (In this study, we focus on cases that receive Ongoing Services depicted in the orange boxes)}
\label{fig:cws_timeline}
\end{figure*}

\subsection{Data Overview} \label{sec:dataoverview}

For this project, we focused on casesnotes written for child welfare cases that received Ongoing Case Management Services (“Ongoing Services”) from the agency. When concerns regarding a child are reported to the agency, and substantiated through initial assessments and investigations, and it is determined that the family requires Ongoing Services to protect the child from future harm, the case is formally transferred to Ongoing Services. In Figure \ref{fig:cws_timeline}, we show a simplified case flow diagram showing how cases are transferred to Ongoing Services (yellow boxes in the Figure).

The first 30 days in receiving Ongoing Services are a critical period where the foundation for subsequent casework decisions is made. During the first month, caseworkers gather and holistically assess detailed information on family functioning, strengths, and unique needs. Caseworkers then work collaboratively with family members to draw up a \textbf{Service Plan} that outlines: (1) objectives, (2) activities, (3) the rationale for each objective and activity, (4) agreed-upon timeframes for completion, and (5) the family’s unique case reference number. Objectives are designed to depict higher-level, longer-term goals for a case while Activities are intended to depict shorter-term, actionable goals that will allow the family to achieve Service Plan Objectives. A short Reason is also often included to provide brief contextual rationale for setting a particular Service Plan's Objectives and Activities. For every Service Plan Objective, there can be more than one corresponding Activity and Reason. Below we show a paraphrased and condensed example of a Service Plan’s Objective, Activity, and Reason.

\begin{myquote}
    \hspace*{2em}\textit{\textbf{Objective:} “For [name] to be raised in a household whose parents demonstrate healthy conflict resolution”}\\
    \hspace*{2em}\textit{\textbf{Activity:} “[name] to acquire effective methods for controlling his anger.”}\\
    \hspace*{2em}\textit{\textbf{Reason:} “The children have observed conflicts between their parents, including physical aggression and objects being thrown.”}
\end{myquote}

The Service Plan is an important document that links caseworker assessments with interventions. Per ministry guidelines, the Plan's goals and reasons are written in standardized and easy-to-understand language because the goal of the document is to provide an actionable framework against which progress can be measured over time for families and workers. After the first month, caseworkers are tasked with meeting the family regularly (at a minimum once a month) and providing services that support the achievement of the Service Plan’s terms. Formally, the Service Plan is reviewed every six months with families, and a new plan may be drawn up reflecting a family’s progress and change in the child or family’s circumstances. 

In addition to the Service Plan, throughout a family’s engagement with the agency, caseworkers record contemporaneous \textbf{Regular casenotes} that document any information pertaining to the case, including detailed observations and email/text/phone/in-person conversation records with families and other service providers such as doctors or lawyers. Each Regular Casenote includes a unique family reference number and the date, time, and location at which the recorded event took place. Documenting casenotes is a critical responsibility for caseworkers, as it informs workers about current efforts made to support children and informs workers on the client’s history \cite{nelson2024designing, geiger2021assessment}. Service Plan Objectives and Activities are closely tied with Regular Casenotes because the Service Plan sets the goals that guide caseworkers’ efforts in supporting families, while Regular Casenotes provide real-time updates on how these goals are being achieved by families and facilitated by workers. Based on observable changes in family functioning and parenting, the agency will close a case when there is no longer evidence of safety threats to the child and improvements have been made in the terms outlined in the most recent Service Plan.

\subsection{Dataset Description}

\begin{minipage}{0.5\textwidth}
\begin{table}[H] 
\Small
\centering      
\begin{tabular}{c | c | c}  
\hline         
\textbf{Percentile} & \textbf{Case Duration (Days)}  & \textbf{\textit{N}}\\ [0.5ex] 
\hline
0-25\% & 5 - 145 & 180  \\   
25-50\%  & 146 - 232 & 180\\ 
50-75\% & 233 - 387 & 181 \\ 
75-100\% & 388- 840  & 179\\ [1ex]
\hline      
\end{tabular}
\caption{Dataset Case Duration Percentiles and Counts} 
\label{tab:duration_percentile}  
\end{table}
\end{minipage}
\hfill
\begin{minipage}{0.5\textwidth}
\begin{figure}[H]
\vspace{-0.2cm}
 \centering
 \includegraphics[width=0.85\textwidth]{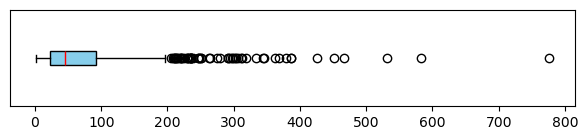} \caption{Distribution of the Number of Regular Casenotes Per Family}
 \label{fig:regcasenote_dist}
\end{figure}
\end{minipage}

\textit{Dataset} For this study, we obtained Regular and Service Plan casenotes from 720 families (i.e., 720 `cases') who received Ongoing Services at the agency between January 2022 and June 2025 and were discharged by June 2025 (henceforth called the `dataset'). Cases ranged in duration with the shortest case being 5 days and the longest being 840 days, averaging 275 days, overall. In Table \ref{tab:duration_percentile}, we show the distribution of case durations for our dataset and a breakdown of the number of families within each duration quartile range. As seen in the Table \ref{tab:duration_percentile}, we found that the agency engaged with an approximately equal number of families within the four duration quartile ranges, suggesting that the agency was handling an equal distribution of cases of varying complexities. Our collaborators in this study expressed a particular interest in understanding why some cases were falling within the 75\% to 100\% percentile of case duration range. 

\textit{Regular Casenotes} Our dataset had a total of 52,748 regular casenote records for the 720 families. On average, a family had a mean number of 73 regular casenote entries but as seen in Figure \ref{fig:regcasenote_dist}, the number of regular casenotes entered for each family varied significantly. 

\textit{Service Plans} Not all families had Service Plans, especially if a case was closed within 30 days. In our dataset, we had a total of 1,213 Service Plans for 654 families, which consisted of a total of 1677 Objectives, and 2,288 Activities. As seen in Table \ref{tab:num_agree_dist}, each family typically only had 1 to 2 Service Plans drawn up during their engagement with the agency although longer cases had multiple Service Plans which were drawn up during bi-annual Service Plan review meetings (see Figure \ref{fig:cws_timeline}). 

\begin{table}[h!]
\centering
\begin{tabular}{cc}
\hline
\textbf{Number of Service Plans} & \textbf{Number of Cases} \\
\hline
1 & 320 \\
2 & 188 \\
3 & 97 \\
4 & 31 \\
5 - 7  & 18 \\
\hline
\end{tabular}
\caption{Distribution of the Number of Service Plans Per Family}
\label{tab:num_agree_dist}
\end{table}

\section{Eliciting Thematic Connections between Regular Casenotes and Service Plan Objectives and Activities (RQ1)} \label{sec:study1}

In Section \ref{sec:dataoverview}, we explained how the goals set in the Service Plan are closely related to Regular casenotes: Service Plan goals guide the delivery of services to clients; Regular casenotes then document family progress in relation to the goals while also capturing dynamic, changing circumstances of clients, which in turn, can reshape Service Plan goals. Considering the close, symbiotic relationship between the two documents, we were interested in identifying thematic similarities and dissimilarities between the two documents in this study. To this end, we applied topic modeling (specifically BERTopic \cite{grootendorst2022bertopic}) on the 1) Service Plan’s Objectives, 2) Activities, and 3) Regular casenotes. In the following subsections, we describe our data preprocessing steps, analysis approach, and findings.

\subsection{Data Cleaning, Preparation, and Anonymization}

For each document type (i.e., Service Plans and Regular casenotes), we extracted and saved core information from the document into a tabular format, including a case’s reference number and date of record. For Service Plans, we also extracted information on the time period for which the goals will be worked on, as well as the Plans’ objectives, activities, and reasons; and for Regular casenotes, we extracted narrative records. 

Following extraction, we cleaned the data, removing extraneous whitespace, line breaks, and special characters. To prevent distinct clustering based on language, we removed French words by passing each sentence through the langdetect library \cite{langdetect}, and removing sentences that have been detected as French. An additional manual cleaning step was taken afterwards to remove any repetitive French sentences that langdetect missed. Following cleaning, we anonymized personal information to protect the privacy of the persons within the cases. The anonymization process was conducted with the following 3 steps. First, we removed personally-identifying information, namely email addresses and phone numbers, replacing them with [email] and [phone], respectively. Next, we replaced client names with the label [client name], so that information pertaining to the client could remain identifiable and not be lost in anonymization. Lastly, we used the spaCy English language model “English transformer pipeline” (en\_core\_web\_trf) to identify words within the texts that correspond to a person and replace that section of text with the label [person]. We used a transformer-based language model to anonymize because, unlike strict string-matching methods, it can identify uncommon names and names that have other meanings as words.

\subsection{Data Analysis Approach}
To generate topics, we ran BERTopic \cite{grootendorst2022bertopic}, then conducted a manual review to determine the types of topics and themes within the casenotes. BERTopic was selected for its ability to extract nuanced semantic information from text with its transformer-based language model embeddings, and for its superior performance in topic coherence and diversity compared to LDA and LMF \cite{grootendorst2022bertopic}. We ran BERTopic using all-MiniLM-L6-v2 embeddings, UMAP, and HDBSCAN. We used UMAP for dimensionality reduction due to its ability to preserve local structure, which facilitates clustering, while also approximately preserving global structure, aiding in the interpretation of semantically similar clusters \cite{mcinnes2018umap}. We used HDBSCAN to accommodate topic clusters of varying sizes. We set a low minimum cluster size to promote classification and reduce the number of unclassified data points, helping to ensure that the generated topic model was representative of the corpus. Hyperparameters of the models were selected based on two criteria: number of topic clusters, and topic coherence. Upon each adjustment to the model hyperparameters, we first examined the number of topics. If there were too many topics to feasibly manually examine, the model was retrained with adjustments. Otherwise, the representative documents of each topic were reviewed to determine if there is a coherent and identifiable theme; if not, the model was retrained. 

For each of the topic models trained on the Objectives, Activities, and Regular Casenotes, four of the co-authors of the paper employed an open-coding process and manually examined documents and keywords assigned to each topic to manually assign a label to each topic and group the topics into high-level topics \cite{braunclark2006}. Two other co-authors of the paper who work at our collaborating child welfare agency then checked and corrected the high-level topics based on their expertise \cite{membercheck2010}. After iterative discussions, we agreed on the high-level topics that emerged in the Services Plan’s Objectives, Activities, and Regular casenotes. Through these interpretive steps, we found the three different narrative types shared similar thematic topics, albeit with some differences.


\subsection{Results: Thematic Connections Between Regular Casenotes and Service Plan Objectives and Activities} \label{sec:r1}

Training BERTopic models on the Regular casenotes generated 60 topics, the Service Plan Objectives yielded 38 topics, and Activities produced 82 topics. Manual inspection of the topics showed that having a high number of topics for each narrative document type allowed for a comprehensive and granular understanding of the wide range of topics covered in the dataset, but these topics could also be grouped into higher-level themes as they contained thematic similarities. As the aim of identifying topics in the Regular Casenotes and Service Plan goals was to understand thematic convergences and divergences between the narrative text types, through iterative and collaborative discussions, we identified high-level themes from the Regular casenotes, Service Plan Objectives, and Service Plan Activities. See Appendix \ref{appendix:example_theme} for examples of how topics were grouped for thematic similarity.

In Table \ref{tab:r1}, we provide an overview of the themes that emerged across the different narrative types, highlighting those that were shared and those unique to each narrative text type. As seen in the table, the three types of narrative texts shared many thematic similarities as the themes point to core child welfare-related issues such as parenting, health, child custody, and safety. At the same time, thematic differences emerged between the three narrative types due to intrinsic differences in the different narrative text types. The following bullet points explain some of the most shared and unique themes that emerged. See Appendix \ref{appendix:super_topic_desc} for exemplar sentences for the below listed themes.

\begin{table*}[]
\resizebox{\textwidth}{!}{%
\begin{tabular}{l|c|c|c}
\multicolumn{1}{c|}{\textbf{Super Theme}} & \textbf{Regular Casenote} & \textbf{Service Plan - Objective} & \textbf{Service Plan - Activity} \\ \hline
Family Relationship/ Visits \& Parenting & x & x & x \\
Child Custody \& Criminal / legal & x & x & x \\
Daycare \& Child Equipment & x & x & x \\
Attempts to Contact & x &  &  \\
Medical/Mental Health & x & x & x \\
Anger Management Conflict \& Safety & x & x & x \\
Housing, Home Environment \& Adoption & x & x & x \\
School & x & x & x \\
Kinship & x & x & x \\
Administration related tasks, resources, and scheduling & x & x & x \\
Support Network &  & x & x \\
Child Development &  & x & \\
\end{tabular}%

}
\vspace{2pt}
\caption{Super themes identified in Regular casenotes, Service Plan Objectives, and Activities (The last three columns indicate whether the theme emerged for the specific narrative text type)}
\label{tab:r1}
\end{table*}

\begin{itemize}
  \item\textbf{(Shared Theme) Family Relationship/ Visits \& Parenting} This theme was about improving relationships between bio-parents and children by promoting healthy family communication, constructive discipline practices, participating in parenting programs, and establishing access visits with parents and children when a child is not placed with the bio-parent. 

\item\textbf{(Shared Theme) Child Custody \& Criminal/Legal}
This theme was centered on topics related to child custody or legal and criminal justice related issues impacting a family. When bio-parents were no longer together, child welfare workers were tasked with connecting parents to mediators who can resolve child custody disagreements and caseworkers also supported family members navigate court proceedings. This theme emerged across all narrative document types.

\item\textbf{(Shared Theme) Medical/Mental Health}
This theme appeared in all goal types and regular casenotes and focused on any medical, health, or substance abuse related issues. The theme could include the facilitation and scheduling of medical appointments, checkups, mental health, and substance abuse treatment appointments as well as families working on developing coping strategies when faced with mental health challenges. 

\item\textbf{(Shared Theme) Anger Management Conflict \& Safety}
This theme appeared across all narrative types and encompassed topics such as child or parent aggression and intimidation, parental conflicts occurring in front of children, and parenting-related safety concerns.

\item\textbf{(Shared Theme) Administration related tasks, resources, and scheduling}
This theme appeared most frequently in the regular casenotes and also emerged in the Objectives and Activities. The focus of this theme was largely on caseworkers facilitating the delivery of child welfare support services to families, interpretation services, referrals, transportation, material resources, and other appointments. Under this theme, caseworkers also conducted administrative tasks, putting together consent forms, obtaining signatures, and providing financial aid to families. 

\item\textbf{(Unique Theme) Attempts to Contact}
This theme only appeared in regular casenotes because it was about caseworkers attempting to reach out to various parties, including family members and other service providers.

\item\textbf{(Unique Theme) Support Network}
This theme focused on workers helping families connect and build support systems so that when challenges arise, families can draw on friends, relatives for support and are knowledgeable of available community resources. This theme did not emerge as a distinct category in the Regular Casenotes, as building a support system typically involved workers connecting clients to community service providers and encouraging parents to engage with friends and relatives. As a result, in Regular casenotes, this theme appeared under other themes such as “Family Relationship/ Visits \& Parenting” and “Daycare \& Equipment” where workers assisted families connect with child supplies and daycare programs.  

\item\textbf{(Unique Theme) Child Development}
The theme of ensuring that a child was developing age-appropriately emerged as a unique category only in the Service Plan Objectives. Similar to the “Support Network” theme, this occurred because achieving this Objective required caseworkers to connect children with schools, counseling services, and health professionals. Consequently, the operationalization of this theme appeared under other themes such as “Family Relationship/Visits \& Parenting,” “Medical/Mental Health,” and “School” in the Regular Casenotes and Service Plan Activities.
\end{itemize}

Upon manual examination of the 38-topic model solution for Service Plan Objectives and the 82-topic model for Activities, we found that high-level goals (i.e., Objectives) in child welfare at the agency are more standardized, while activities tend to be more customized to the case. Activities provided concrete, actionable steps to achieve a case's Objective. Guided by the Service Plan goals, the narrative information in Regular casenotes directly reflected how caseworkers supported their clients guided by pre-established Service Plan goals while also adapting to clients' unique circumstances. See Appendix~\ref{appendix:obj_act_desc} for a more in-depth analysis and distinctions between Service Plan Objectives and Activities.

\section{Mapping Activity-relevance in Regular Casenotes (RQ2)} \label{sec:study2}

In the second part of this study, we sought to examine how computational text analysis approaches (i.e., a LocalLLM) could be used to track Service Plan goal progress in Regular casenotes. We employed a LocalLLM because they are open-source LLMs that run in offline environments and do not upload any data to the cloud, unlike their cloud-based counterparts. The local deployment of LocalLLMs ensures the protection and full control of sensitive private data, while still offering sufficient computational capabilities of modern LLMs \cite{perron25}. Our results from Section \ref{sec:study1} revealed that Activities provided actionable guides for families to achieve overarching Service Plan Objectives. We hypothesized that Activities could better track case progress than Objectives, so we employed a LocalLLM to label which Regular casenotes mentioned Service Plan Activities were being completed or worked on and manually reviewed the labels.

\subsection{Data Analysis Approach} \label{sec:sec6.1}

Llama 3.1 \cite{llama31} was selected for the LocalLLM data analysis tasks. This model was selected for its lightweight, being 4.9GB in size and having 8 billion model parameters, allowing it to be stored in system RAM on inference on local devices, and for its superior run-time performance compared to state-of-the-art 7-10 billion parameter lightweight LLMs of similar size such as Mistral, PHI4 and deepseek-R1 \cite{grattafiori2024llama}. Llama 3.1 was run on the Ollama platform \cite{ollama} due to its ability to work on multiple operating systems and devices.


To track Service Plan Activity progress within Regular casenotes, we sought to directly match Activities to Regular casenotes by including both the Activity and the casenote within the prompt, and querying if within that Regular casenote, if there is any indication of progress towards the Activity. To mitigate potential issues around prompt brittleness, where small changes in prompt formats can lead to inconsistent performance fluctuations \cite{promptbrittleness}, we experimented with multiple prompt formats and selected the prompt below, which generated responses most accurately and relevant to our needs. Specifically, to reduce the varying structures and formats of the text from impacting the Activity-casenote tracking, the LocalLLM was tasked with first generating a summary of the contents of the casenote, and using that generated summary to see if there is indication of progress. The length of the summary remained unspecified, allowing for longer summaries to be generated for content rich casenotes, helping preserve information. We provide the prompt below:

\begin{myquote}
    \hspace*{1em}\texttt{\textit{\textbf{Activity-Regular Casenote progress tracking prompt:}} \\
    \hspace*{1em}\texttt{You are analyzing a child welfare worker reviewing case notes. Do the following:} \\
    \hspace*{1em}\texttt{1. Read through the casenote and store it in summary.}\\
    \hspace*{1em}\texttt{2. Assess whether the summary indicates progress toward completing or working on the following} \\
    \hspace*{1em}\texttt{activity: \{activity\_name\}. Answer strictly `Yes' or `No'.}\\
    \hspace*{1em}\texttt{Case Note: \{narrative\_text\}}}
\end{myquote}

We ran the LocalLLM for a subset of 100 cases (out of the 720 cases in our dataset) where cases had Regular casenotes and at minimum, at least one Service Plan. To ensure we had a representative sample of cases, we randomly selected 25 cases within each case duration percentile as depicted in Table \ref{tab:duration_percentile}. Our subset of 100 cases included 25 cases that fall within the 5-145 days range ("Short Cases"), 25 cases that fall within the 146-232 days range ("Medium Cases"), 25 cases that fall within the 233-387 days range ("Long Cases"), and 25 cases that fall within the 388+ days range ("Extreme Cases"). The first author also manually read through each regular casenote entry for the 100 cases (\textit{N}=6,031) to compare the accuracy of the LocalLLM’s labels with a human labeler and resolved ambiguous casenotes by consulting with two co-authors of the paper who work at the agency. In Table \ref{tab:progresstracking_descstats}, we show the average number of words within each Regular casenote entry for cases of different duration types for our subset of 100 cases. As seen in the table, all cases, irrespective of case length, had approximately similar number of words in each casenote.

\begin{table*}[]
\resizebox{\textwidth}{!}{%
\begin{tabular}{l|ccc|cc}
 & \multicolumn{3}{c|}{\textbf{Average Number of Words}} & \multicolumn{2}{c}{\textbf{Total Count}} \\ \cline{2-6} 
\textbf{Case Duration} & \multicolumn{1}{c|}{\textbf{Regular Casenote Entry}} & \multicolumn{1}{c|}{\textbf{Service Plan - Objectives}} & \textbf{Service Plan - Activities} & \multicolumn{1}{c|}{\textbf{Regular Casenote Entry}} & \textbf{Service Plans} \\ \hline
\textbf{Short} & \multicolumn{1}{c|}{230.1} & \multicolumn{1}{c|}{9.44} & 13.57 & \multicolumn{1}{c|}{530} & 27 \\
\textbf{Medium} & \multicolumn{1}{c|}{256.6} & \multicolumn{1}{c|}{10.79} & 12.59 & \multicolumn{1}{c|}{925} & 26 \\
\textbf{Long} & \multicolumn{1}{c|}{279.5} & \multicolumn{1}{c|}{9.41} & 12.52 & \multicolumn{1}{c|}{1485} & 42 \\
\textbf{Extreme} & \multicolumn{1}{c|}{241.0} & \multicolumn{1}{c|}{9.38} & 12.14 & \multicolumn{1}{c|}{3095} & 76 \\
\textbf{All cases} & \multicolumn{1}{c|}{251.9} & \multicolumn{1}{c|}{9.63} & 12.53 & \multicolumn{1}{c|}{6031} & 171
\end{tabular}%
}
\vspace{2pt}
\caption{Average number of words and total counts of the cleaned and anonymized Regular casenotes and Service Plans for the subset of 100 cases for progress tracking. }
\label{tab:progresstracking_descstats}
\end{table*}

\subsection{Results}

\subsubsection{\textbf{Manually labeling shows approximately half of the Regular casenotes mentioned Activity relevant information except for Extremely long cases}} \label{sec:prop_relevance}

\begin{figure*}[]
\centering 
\includegraphics[scale=0.4]{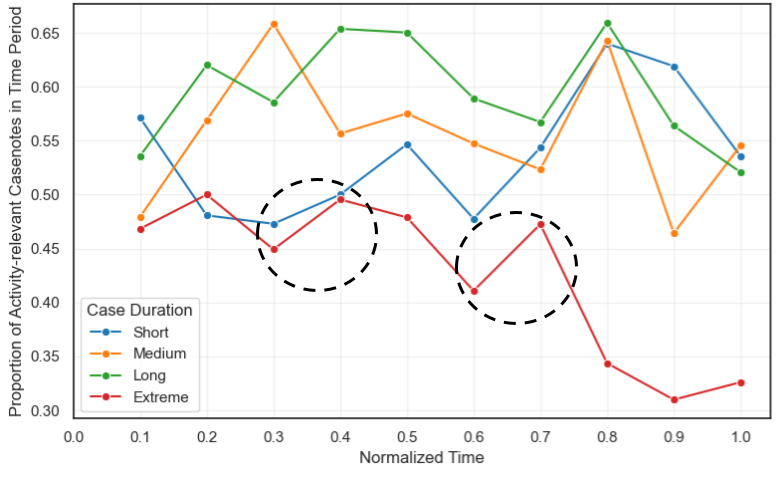}
\caption{Proportion of Regular casenotes mentioning Activity-relevant information for each normalized time period. New Service Plans are typically drawn up during the 0.3 and 0.6 time period for Extremely long cases (denoted in black dotted circles).}
\label{fig:exp3}
\end{figure*}

In Figure \ref{fig:exp3}, we show the proportion of regular casenotes that mention Activity-relevant information following our manual labeling. Along the x-axis, we show a normalized timeline of a case. Prior work by Saxena et al. \cite{saxena22} demonstrated that child welfare casenotes follow a structured sequence of events, and that segmenting cases into equal intervals can reveal temporal trends across cases of varying durations, illustrating the `Life of a Case’. We adopted this methodology to depict temporal trends in cases such that if a case duration was 100 days, casenotes were segmented such that regular casenotes written during the first 10 days fell within the 0.1 normalized time period. On the y-axis, we show the proportion of regular casenote entries that mention Activity-relevant information for the specific normalized time period. As shown in the figure, across each normalized time period, Short, Medium, and Long cases generally included Activity-relevant information in approximately 46–66\% of the Regular casenotes written for each period. However, Activity-relevance consistently dropped towards the latter half of Extremely long cases. Manual readings of the casenotes revealed that there were decreases in the proportion of Activity-relevant regular casenotes in Extremely long cases when new child welfare concerns not outlined in the Activities emerged. For example, a family Activity could be set to be \textit{"[name] to engage in positive communication and interactions with [name]"}. However, during the timeframe in which the Activity must be completed, a bio-parent may be placed under house arrest and unable to leave home to meet their child. In this case, regular casenotes would no longer include information related to the Activity. We only observed slight increases in Activity-relevance in casenotes during the 0.4 and 0.7 time period when new Service Plans would be typically drawn up for Extreme cases. New Service Plans would address the updated needs of families but even then the rebounds in Activity-relevance were shortlived for complex cases (see the round dotted circles in the Figure). In the following paragraphs, we compare how LocalLLMs labeled Activity-relevance in casenotes compared to manual labels. 

\subsubsection{\textbf{Local LLMs can Indicate Case Progress and Deviations in Less Complex, Shorter Duration Cases}}

In Table \ref{tab:exp3} we present inter-rater reliability metrics comparing our manual labels with LocalLLM-generated labels for identifying Service Plan Activity content in regular casenotes. As shown in the Table, Short-length cases (5-145 days) achieved a higher Cohen's kappa of 0.604, but as cases became longer in duration, the kappa coefficient decreased to 0.402 for Long cases (233-387 days) and 0.470 for Extremely long cases (388-840 days). The false positive rate (FPR) slightly increased as cases became longer in duration, while the false negative rate (FNR) remained relatively flat across case durations. This suggested that the LocalLLM tended to label regular casenote entries as Activity-relevant and could not correctly identify Activity-relevant narratives for more longer duration child welfare cases. Manual readings of the cases showed that cases that fell within the Long and Extremely long duration cases involved more topics related to access visit coordination, as children were often placed outside of the home and frequently involved court, legal issues, and more referrals to external service providers. Compared to Short and Medium cases, we observed that underlying issues that initially brought a family to the agency often escalated and triggered a chain of events (often unspecified in Service Plan Activities) that prolonged a family's engagement with the agency in Long and Extreme duration cases, which were then often incorrectly labeled as Activity-relevant by the LocalLLM.

\begin{table*}[]
\begin{tabular}{l|c|c|c|c}
\textbf{Case Duration} & \multicolumn{1}{l|}{\textbf{Agreement}} & \multicolumn{1}{l|}{\textbf{Cohen's $\kappa$}} & \multicolumn{1}{l|}{\textbf{FPR}} & \multicolumn{1}{l}{\textbf{FNR}}  \\ \hline
\textbf{Short} & 0.804 & 0.604 [0.535, 0.670] & 0.265 & 0.135 \\
\textbf{Medium} & 0.782 & 0.550 [0.493, 0.602]& 0.324 & 0.135  \\
\textbf{Long} & 0.727 & 0.402 [0.352, 0.447] & 0.470 & 0.146  \\
\textbf{Extreme} & 0.730 & 0.470 [0.440, 0.500] & 0.382 & 0.134 \\
\end{tabular}
\vspace{2pt}
\caption{Inter-rater reliability between manual and LocalLLM progress tracking, with 95\% bootstrapped confidence intervals, and LocalLLM classification performance by case duration category }
\label{tab:exp3}
\end{table*}

Comparisons between manual and LocalLLM labels showed that the LocalLLM's tendency to predict regular casenote entries as Activity-relevant could be attributed to the model's prompt configuration limitations and dataset characteristics. For example, Activities often used language that could appear vague to the lay reader, but had specific meanings in child welfare literature. If an Activity stated, \textit{"Parents to develop a safety plan in case [name] returns to family home"} or \textit{"The family will work toward enhancing their support system,"} the LocalLLM would identify any texts related to general safety or support as Activity-relevant. Furthermore, due to the highly sensitive nature of the data, we had anonymized all identifying information in the casenotes. This meant that when an Activity called on a specific person to carry out an action, the LocalLLM labelled any casenote that mentions any person conducting the action as Activity-relevant. When we manually labeled the texts, we could generally infer which bio-parent the Acitivity was relevant for based on their pronouns and infer who was carrying out the actions in the casenotes based on contextual details. 

Table \ref{tab:exp3} shows that the LocalLLM produced approximately 13-14\% false negative errors. We observed that these errors often occurred when the LocalLLM could not infer how acronyms or specific service provider names were related to an activity. For example, an Activity may state, \textit{"[name] to engage in the recommended substance use/mental health related services and follow through with recommendations."} If a Regular casenote entry said, \textit{"[name] from [hospital name acronym] advised that [name] is currently with them and that she came in on the 17th and that she was put on a Form 1 and that she is currently staying voluntarily,"} the LocalLLM could not understand that the hospital acronym and specific mental-health related terms (i.e., Form 1) could indicate Activity-relevance. 

When manually comparing LocalLLM labels with our manual labels, we also considered whether prompt brittleness could be leading to the misclassifications of Activity-relevance \cite{promptbrittleness}. In most cases, however, we could trace the false positive and false negative errors back to the reasons detailed above, and to underlying ambiguities in how case progress is defined, which we elaborate on below.

\subsubsection{\textbf{Ambiguities arise on what is considered relevant Activity-relevant and require discretionary judgments grounded in social work training}} \label{sec:r2_ambiguity}

It is important to emphasize that the exercise of using a LocalLLM to track Activity progress was not to build a perfect and deployable progress tracker for workers. Instead, the aim of employing a LocalLLM and manually labeling regular casenotes was to uncover challenges and opportunities for applying LocalLLMs within the child welfare domain. Critical limitations to applying LocalLLMs to track Activity progress surfaced when there were casenotes that the first author could not easily label as Activity-relevant and required consultations with two co-authors of the paper who work at the agency to draw on their social work expertise. For example, there was one particular case where the biological father and biological mother were separated and often at odds with each other. An Activity for the family was, \textit{“The society will engage the parents in working with a family mediator to resolve custody and access arrangements.”} Manual reading of the casenotes showed that due to parental conflict, the caseworker unwillingly acted as a mediator for the bio-parents to schedule visits with the children, even though parental mediation is not part of their main job duties. In this case, it was unclear if the caseworker acting as an (unwilling) mediator would be considered Activity-relevant. 
To see a more full picture of the diversity of ambiguous cases encountered, see Appendix~\ref{appendix:case_ambg}.

Discussions to resolve Activity-relevance ambiguities with our co-authors from the agency demonstrated that determining which casenotes are relevant to a Service Plan’s Activities requires discretionary judgments grounded in social work training and is not a straightforward task. Instead, temporal and contextual factors surrounding a case must be considered, which can be particularly challenging to parse, define, and operationalize through LLM prompting. 

\section{Thematic Tracking of Activity-Relevant Regular Casenotes (RQ2)} \label{sec:exp4}

Our findings from Section \ref{sec:study2} indicated which Regular casenote entries contained Activity-relevant information but did not provide information on the narrative content of the texts. To delve into content-specific, thematic trends within the casenotes, we built on the Super Themes identified in Table \ref{tab:r1} in Section \ref{sec:r1} and applied a LocalLLM to compare thematic narrative trends in Regular casenote entries that had been manually labeled as Activity-relevant and Activity-irrelevant. 


\subsection{Data Analysis Approach}

To track the thematic content of Regular casenotes, we prompted a LocalLLM to identify themes that are present in each Regular casenote entry. We provided the LocalLLM a list of Themes, drawing on the Super Themes we had identified to be present in the Regular casenotes (see the second column Table \ref{tab:r1}). We used these themes because all of the Activities Super Themes are also present within these set of themes, all except for "Support Network", whose corresponding activities can be classified under the other themes such as \textit{"Family Relationship/ Visits \& Parenting"}, \textit{"Daycare \& Equipment"}, and \textit{"Administration related tasks, resources, and scheduling"}. By applying this approach, we could determine the themes of focus within each Regular casenote entry. Below, we provide a segment of the LocalLLM prompt. The prompt details the instructions for matching and the preset list of themes that the Regular casenotes should look out for. Similar to how our prompt was developed for Section \ref{sec:sec6.1}, we adopted an iterative approach, experimenting with different prompt formats to mitigate potential prompt brittleness issues.

\begin{myquote}
{\ttfamily
    \hspace*{1em}\textit{\textbf{Regular Casenote Theme extraction prompt:} \\
    \hspace*{1em}You are a social work assistant working in a child welfare agency. Your task is to identify within the casenote whether or not a provided theme is mentioned.\\
    \hspace*{1em}Indications must be: Match or No Match. \\
    \hspace*{1em}- "Match" means that the activity content falls within the theme or is directly related to it. \\
    \hspace*{1em}- "No Match" means the activity is not relevant to the theme.\\
    \hspace*{1em}casenote: \{casenote\} \\
    \hspace*{1em}Themes (with descriptions): \\
    \hspace*{1em}1. Family Relationship/ Visits \& Parenting \\
    \hspace*{1em}Description: improving parent child relationship, such as communication in the family and discipline practices, parenting programs and establishing access and visits with parents and family members \\ \\
    \hspace*{1em}2. Child Custody \& Criminal / legal\\
    \hspace*{1em}Description: figuring out custody over child and or going to court to deal with criminal legal... } \\
    }
    
\end{myquote}

Model generation parameters were adjusted to provide more consistent generated outputs, including adjusting the temperature parameter to a low value of 0.1, and specifying the json output format. To assess the reliability of the LocalLLM to correctly identify themes in the texts, two of the co-authors manually spot checked some of the thematic labels. 

\subsection{Results: LocalLLMs can Surface Thematic Divergences in Regular Casenotes}

In this section, for improved readability, we present thematic narrative trends for Short and Extremely long cases for four Super Themes that are central to child welfare concerns. 

\begin{figure*}[t]
\centering 
\includegraphics[scale=0.5]{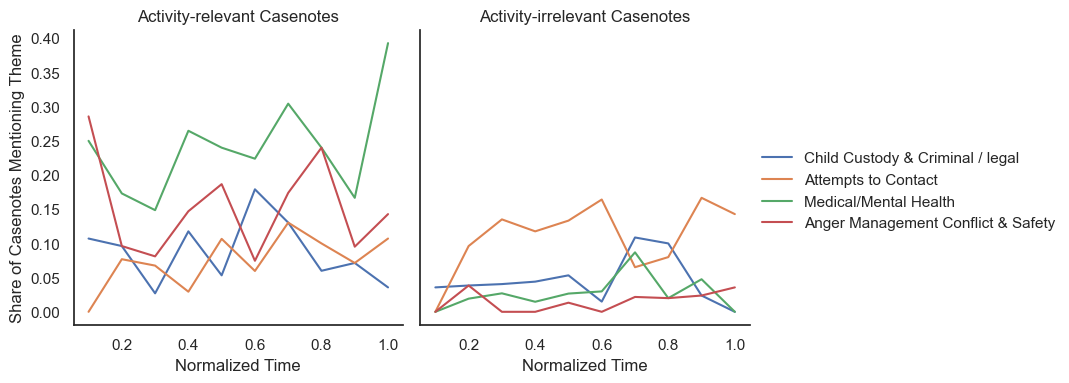}
\caption{Proportion of Themes present in Regular Casenotes for each normalized time period for Short Duration Cases}
\label{fig:exp4_short}
\end{figure*}

The facet plots in Figure \ref{fig:exp4_short} show the proportion of Regular casenotes mentioning Super Themes across normalized time periods for Short cases, separated by casenotes manually labeled as Activity-relevant (figure on left) and Activity-irrelevant (figure on right). For example, the figure on the left shows that of all the Short-duration regular casenotes written during the 0.1 time-period, approximately 25\% of Regular casenotes contained information that was 1) Service Plan Activity-relevant and 2) relating to the \textit{Medical/Mental Health} theme. 

From the left facet plot of Figure \ref{fig:exp4_short}, we observe stepwise increases in the four Super Themes throughout the life of a case. This suggests workers generally closely follow Service Plan Activities in Short cases. The left line plot also shows that Activity-relevant casenotes that mentioned \textit{Medical/Mental Health}, \textit{Anger Management Conflict \& Safety}, and \textit{Attempts to Contact} themes become more prominent at the final stage of a case, i.e., at normalized time =1.0. Through manual reading of the casenotes, we discovered this increase occurred at the end of cases because workers often contacted clients to inform them that the agency was closing the case due to a lack of safety concerns.

From the facet plot on the right in Figure \ref{fig:exp4_short}, which shows the proportion of regular casenotes that do not contain Activity-relevant information for each time period, we observe comparatively fewer mentions of the four Super Themes. We observe that workers support clients on issues related to \textit{Child Custody \& Criminal/Legal matters} and \textit{Medical/Mental Health} that are not relevant to Service Plan Activities during the latter half of a case, but their prominence decreases by the end of the life of a case. Manual reading of casenotes showed that, because Short-duration cases were often relatively straightforward, workers could quickly support clients in these areas by connecting them to available service providers.

\begin{figure*}[t]
\centering 
\includegraphics[scale=0.5]{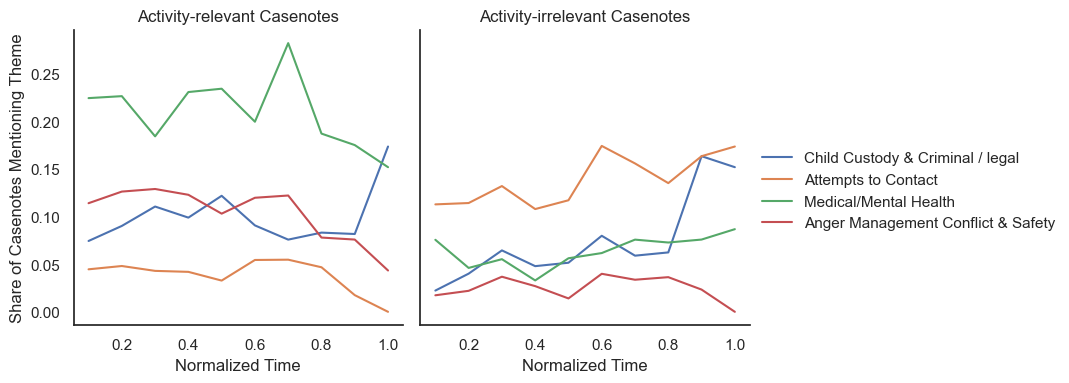}
\caption{Proportion of Themes present in Regular Casenotes for each normalized time period for Extreme Duration Cases}
\label{fig:exp4_extreme}
\end{figure*}

Compared to Short cases, we observe different thematic trends for Extremely long cases. Figure \ref{fig:exp4_extreme} shows the proportion of Regular casenotes that mention Super Themes across normalized time periods for Extremely long cases, separated by casenotes manually identified as Activity-relevant (left) and Activity-irrelevant (right). From the left facet plot in Figure \ref{fig:exp4_extreme}, we observe that Regular casenotes increasingly diverge from Service Plan Activities over the life of a case. For example, the prominence of casenotes that are Activity-relevant and mention themes, \textit{Medical/Mental Health} and \textit{Anger Management Conflict \& Safety} decrease steadily over the life of a case, only temporarily rebounding during the 0.7 time period when new Service Plans are often drawn up. We also note that workers are making fewer \textit{Attempts to Contact} clients/other related parties regarding Activity-relevant issues (Figure \ref{fig:exp4_extreme} left plot), and instead, workers are making more \textit{Attempts to Contact} people regarding Activity-irrelevant issues (Figure \ref{fig:exp4_extreme} right plot) over time. 

In the right facet plot of Figure \ref{fig:exp4_extreme}, the increases we observe for the four Activity-irrelevant themes suggest that in Extremely long cases, new child welfare concerns, such as those related to a client's health, substance abuse usage, and criminal/legal issues, become more prominent over time and thus renders existing Service Plan Activities outdated due to emergent child welfare concerns. In fact, through manual readings of casenotes, we confirmed this often occurred in Extreme-duration cases. New concerns relating to a bio-parent's substance abuse issues or being implicated in legal or criminal-related matters would emerge, meaning that existing Activities centered on improving child-parent relationships could not be worked on. Instead, custody arrangements had to be made to establish a permanency plan for the child.

 \section{Discussion}

 In this section, we discuss our key findings in relation to our research questions and extend our findings to broader implications for designing LLMs for the public sector. 

\subsection{Key Findings from our Results (RQ1 \& RQ2)} 
In this section, we discuss our key findings concerning RQ1 and RQ2 to reflect on the opportunities and challenges of adopting LLMs for the public sector.

\subsubsection{Promises of Applying LLMs in the Public Sector} The methodology we employed to track thematic trajectories of child welfare cases from Sections \ref{sec:study1} and \ref{sec:exp4} suggests that \textbf{LLMs can provide caseworkers with tools to track narrative content of casenotes with a high level of granularity and customization \cite{perron25},} potentially supporting their clinical decision-making processes. In machine learning literature, prior to the induction of LLMs, traditional inferential statistical methods (such as generalized linear models) and topic modeling techniques have been useful in providing trends on average outcomes and aggregated perspectives \cite{howwedo2020, mlforsociology, grootendorst2022bertopic, lda_blei} but they have also exhibited poor performance against outliers \cite{stevens1984outliers}. In high-stakes domains such as child welfare, effectively adjudicating child welfare risks while avoiding misclassification of low- and high-risk cases, including rare but severe instances of abuse or neglect (i.e., outlier cases), is critical. By combining thematic clusters from BERTopic models with LLMs, our study provides potential opportunities for workers to use computational text analysis techniques as decision-aid tools, enabling birds-eye, exploratory content analysis of large volumes of unstructured text and then narrowing down into specific topics of interest that could signal critical child welfare concerns. This approach could allow workers to identify specific thematic child welfare concerns and opportunistically address gaps in their work to better meet the needs of children and youth. Recent HCI scholarship has argued for moving away from a predictive risk-focused paradigm in child welfare tools built on reductive administrative data, noting that caseworkers view social work as a practice centered on generating knowledge about clients, with day-to-day interactions being central to their professional practice, and that they desire tools that support their knowledge-making practice on clients \cite{Lehtiniemi_2024, saxena2021framework2}. Through our work, we show that nascent language models (such as BERTopic and LLMs) can surface contextualized client information that may help workers make better-informed child-welfare decisions.

\subsubsection{Perils of Applying LLMs in the Public Sector.} \label{sec:perils}
Despite the abovementioned opportunities, our findings also point to the dangers of applying LLMs blindly in domains such as child welfare, where there is high uncertainty and often no clear ground truths. Our findings in Section \ref{sec:study2} show that tracking progress in child welfare cases is not straightforward due to 1) intrinsic uncertainties that underpin child welfare cases \cite{chi23paper} and 2) mixed views between frontline workers and agency management on what constitutes a caseworker’s duties. In this study, we tracked case progress by identifying which Regular casenotes made relevant references to Service Plan Activity goals manually and with a LocalLLM. \textbf{Comparing label agreements between the manual labels and the model showed that the LocalLLM performed increasingly worse for more complex, longer-duration cases, not only due to technical configuration limitations of the LocalLLM prompt\footnote{These limitations could be improved through further prompt engineering or incorporating domain knowledge through retrieval-augmented generation technology \cite{perron_rag2025}.}} \textbf{but also because of fundamental ambiguities on what is considered an Activity-relevant casenote entry}. As noted in Section \ref{sec:r2_ambiguity}, co-authors of the paper, including authors who work at the agency, needed to collaboratively discuss to reach a consensus on whether some casenotes would be deemed Activity-relevant. In cases, where caseworkers took on job roles that were outside of their job duties but were tangentially related to Service Plan Activity goals (e.g., when an Activity was for a caseworker to connect a family with a mediator but the worker took on the role of a mediator themselves albeit unwillingly), our co-authors from the agency argued that from a management perspective, such casenotes may be considered Activity-irrelevant. However, the co-authors also acknowledged that caseworkers often support families in ways that fall outside their official job duties to build rapport with clients, and therefore, could also be considered Activity-relevant \cite{saxena22}. Our agency co-authors explained that child welfare work involves balancing the dual objectives of reducing harm and preserving the family unit \cite{Fluke_dual2016}, meaning interpretations of Activity-relevance can vary between agency management staff and frontline caseworkers, as well as among frontline workers themselves. 

Our findings in Section \ref{sec:prop_relevance} also show that child welfare case goals are not a static construct, and tracking case progress based on documented Service Plan goals can be problematic. In Section \ref{sec:study2}, we found that over a life of a case, Regular casenotes made fewer mentions of activity-relevant content in Extremely long cases and through thematic tracking of casenotes in Section \ref{sec:exp4}, we found this was because caseworkers were addressing new emergent and pressing child welfare concerns that rendered existing Service Plan goals outdated. Our study shows that, similar to how caseworkers can adaptively adjust to the new needs of clients before new and updated Service Plans are drawn up for the family, LocalLLMs would need to draw on up-to-date child welfare goals to be able to track a family’s progress in addressing child welfare-related concerns.

\subsubsection{Practitioner-Grounded Methodological Takeaways for Building LLM-driven Decision-Supports} \label{sec:method_reflection} 

Our study’s methodological steps illustrate how researchers can examine the ways language models might support social-work decision-making. Through participatory collaborations with public sector practitioners, the impetus for this study was motivated by the child welfare agency's commitment to improving decision-making processes. Our methodological decisions to categorize cases based on distributional patterns of a family's average engagement with the agency (i.e., short, medium, long, and extremely long cases) was guided by the agency's operational interests; we defined case progress in accordance with the provincial child welfare ministry's policy guidelines (as recorded in Service Plans); and traced case journeys through the documented lens of frontline workers. Through these research steps, our comparison of the LocalLLM's ability to identify progress against manual labels assessed how well the tools capture progress as articulated by the  
workers' documentation practices, which themselves are shaped by organizational mandates and provincial child welfare policies. Inspired by Antoniak et al. \cite{antoniak19} and Saxena et al. \cite{saxena22}, we also showcase a portable method to compare case trajectories over normalized time periods for different types of cases (Section \ref{sec:prop_relevance}).

\subsection{The Role of LLMs in the Public Sector (RQ2)}

\subsubsection{LLMs as Diagnostic Tools that Center Human Discretion} \label{sec:diagnostictool}

In our study, we find LLMs can be useful for public sector practitioners as diagnostic tools to elicit case trajectories and relevant information for workers \cite{abebe_roles20}. We also note the fallibility of ascribing adjudicative authority to LLMs when determining progress-relevance in documents. As seen from the decreasing Kappa coefficient in our results in Table \ref{tab:exp3} in Section \ref{sec:study2}, our LocalLLM could not correctly identify Activity-relevant casenotes in complex, longer-duration cases because Activity-relevance is dependent on temporally changing clients' circumstances and because what child welfare professionals consider to be Activity-relevant is based on contextual interpretations of a client's ecological environment \cite{saxena2021framework2, Ferguson2009-ng}. In the LLM research space, one could classify the work of child welfare caseworkers as subjective decision-making \cite{paula25}; however, that would be an inaccurate description. \textbf{The crux of high-stakes public sector service work is that work in this space is inherently uncertain. The role of social workers is to 1) gauge and navigate these uncertainties based on best practices in social work scholarship and 2) follow established regulatory practices so that decision-making in this space is defensible and transparent \cite{geiger2021assessment, saxena22, chelmis_21}.} In the case of child welfare, prior work by Saxena et al. \cite{chi23paper} showed that a confluence of systemic, procedural, and risk factors often underpin complex child welfare cases, which can confound caseworkers' decision-making, and these uncertainties can persist even after case closure. Traditionally, when such ambiguities arise, child welfare workers have engaged in collaborative meetings with other caseworkers (including their supervisors) to holistically evaluate a family's ecological environment following child protection regulatory guidelines \cite{ saxena2021framework2, Ferguson2009-ng}. \textbf{Our findings show that LLMs, on their own, cannot and do not replicate child welfare caseworkers' iterative, discursive, and procedural decision-making processes.} And yet, we also note that our practitioner co-authors found value in the LLMs’ potential as decision-support tools for inferring important child welfare signals and case trajectories from documents, particularly compared to contemporary predictive risk model that generate point predictions \cite{moon24, kelly23, showkat23, saxena2020human, mcconvey2024designing}. Therefore, we argue any bureaucratic decision-support tools that employ LLMs should be required to center human discretion (i.e., specialized expertise, value judgements, and heuristic reasoning) \cite{saxena2021framework2} whether that involves human labeling or oversight around the design and use of these tools.

\subsubsection{Designing Public-Sector LLMs Around Local Governance Practices}
Our findings underscore the importance of prioritizing and ensuring effective local governance when utilizing LLMs in the public sector. Currently, most welfare systems lack the resources to develop AI tools in-house and thus often acquire them from private companies \cite{Levy_2021}. Procuring AI technologies can result in power imbalances between public sector agencies and companies. Restrictive procurement contracts can make it difficult for workers to seek information about the model, and workers may come to over-rely on AI technologies, leading to private interests bleeding into public governance practices \cite{redden2020,Kawakami_2024}. These imbalances may be particularly severe for public administrations that lack infrastructural support to accommodate community and practitioner engagement in the AI design/deployment process. To mitigate the private sector co-opting public sector AI governance, we argue 1) local data stewardship should be prioritized by ensuring LLMs can be run locally; 2) impacted stakeholders and practitioners should be actively engaged with the AI development process early on \cite{ Kawakami_2024}; and 3) the onus of justifying the impact and harms that arise from the AI tool should fall on developers \cite{Wang2024}. 

We argue that ensuring local governance first practices is particularly critical in the adoption of LLMs for the public sector because these tools may be vulnerable to many of the same limitations as contemporary predictive risk algorithms, namely 1) the reductive representation of complex phenomena and 2) focus on individualizing societal issues while ignoring systemic constraints \cite{redden_predictive20, keddell15, Mead_Thurston_Bloyce_2022}. The same neoliberal austerity policies that gave rise to public sector algorithmic tools are driving the adoption of LLMs in public agencies, i.e., the desire to do more with fewer resources. For example, the US Department of Homeland Security's pilot projects that will apply LLMs in public agencies state these tools will promote efficiency, and improve the quality of public service work with less resources \cite{dhs_llm}. A similar rhetoric appears in public statements from the Canadian federal government when announcing its partnership with Cohere, a Canadian LLM developer \cite{cohere}. If austerity policies drive the underlying impetus to apply LLMs, these tools could simply enable public agencies to better control access to public services for those deemed eligible, while overlooking individuals who fall through policy gaps and justifying further cuts to essential public programs \cite{eubanks2018automating}. While we acknowledge LLMs hold remarkable potential to transform bureaucratic decision-making, we contend that recent interest in LLMs give rise to one of the oldest problems in HCI research. As Voida et al. \cite{Voida2014} articulated over a decade ago, "even when core values align, tensions may still exist about how to achieve desired ends, or what these values mean in practice." Everyone likely wants to improve public service delivery, however how to do so, we argue, requires further inquiry through collaborations between practitioners and researchers in HCI, computer science, critical data studies, social work, public policy, and beyond.

\subsection{Implications for HCI Research in the Public Sector (RQ3)}

In this section, we propose recommendations for how HCI researchers can engage with public sector practitioners to design LLMs that support their work.

\subsubsection{Building Ground-Up Participatory Partnerships with Public Agencies} \label{sec:rq3_pt1}

We argue that \textbf{HCI researchers should work directly with public sector agencies on issues of key relevance for the organization}. The SIGCHI community has a long history of engaging in participatory methods to promote sustainable human-AI partnerships and supporting the responsible use and deployment of AI tools within the public sector \cite{mothilal25_assumptions, haque24, chui25_compass,weitz24, seyun24, seyun24_2, jo25, tang_failurecards24}. However, oftentimes these works have been conducted through short-term collaborations such as through design workshops that discuss hypothetical AI-related scenarios and interview studies to elicit stakeholder preferences and perceptions \cite{kotturi24, cooper22}. While these studies are useful for understanding how AI tools should be used and designed, they are often extractive and do not necessarily lead to meaningful change for public sector stakeholders \cite{birhane_parti}. There is a greater need for HCI scholars to collaborate directly with public sector organizations to provide critical perspectives on how to design LLM-driven tools \cite{aragon2022human, hcai, to_23, Ogburu20}, particularly because agencies cannot build LLMs in-house \cite{Levy_2021}. Due to the substantial computational resources, energy, and expertise required to build LLMs, most LLMs are built on foundational models developed by a handful of technology companies (in our study, we likewise relied on an off-the-shelf model, Llama 3.1, developed by Meta AI \cite{livebench}). As North American federal governments aggressively seek to adopt AI tools to increase governmental efficiencies amidst austerity measures and techno-optimism, there lies a real concern that government services may be shaped by the interests of big technology companies as government workers increasingly rely on these tools \cite{ redden2020, brookings}. As such, there is renewed urgency for HCI scholars to collaborate directly with public sector organizations to provide critical perspectives on how to design AI tools that are centered on human experiences and needs \cite{aragon2022human, hcai, to_23, Ogburu20}.

We also argue that \textbf{HCI researchers should strive to engage in sustainable long-term community-based research \cite{cooper22, kotturi24} as short-term partnerships may be insufficient for fostering meaningful social change}. In this study, academics and child welfare professionals came together to investigate a question of interest for a large child welfare agency. Through our study, we showed that current off-the-shelf tools are unable to correctly identify progress-relevance in Regular casenotes for complex cases, which are precisely areas in which caseworkers would want the most support. We also realized that our findings generated further research directions that required further examination. For example, our agency co-authors thought it particularly interesting that Extremely long cases had fewer Activity-relevant casenote entries towards the latter half of cases because new issues unrelated to Service Plan Activities emerged (Sections \ref{sec:study2} and \ref{sec:exp4}). Our co-authors reflected that the longer a family remains involved with the agency, the more likely it is that new child welfare concerns will arise. They suggested that future work could assess whether such cases should have been closed earlier or whether families required further services, and explore whether LLMs could detect Service Plan deviations in Regular casenotes to prompt workers to revisit case goals and assess their own practice. Moreover, our agency co-authors noted that for our study’s methodological approach to generate actionable insights, the Super Themes of Table \ref{tab:r1} needed to be broken down further and tracked using the methodology of Section \ref{sec:exp4}. For example, currently, we have grouped substance abuse and counseling-related topics within the \textit{“Medical/Mental Health”} theme (Section \ref{sec:study1}), but agency workers reflected that these two topics could be separated out because substance abuse involves varying levels of need and counseling encompasses a wide range of services. HCI scholarship has long emphasized the importance of empowering communities in participatory research by placing them in the driver’s seat \cite{kotturi24, cooper22, Costanza-Chock_2020}. We provide an example of this approach and highlight the need for long-term collaborations to address substantive research questions of social relevance.

\subsubsection{Operationalizing Ground-Up Partnerships for Public-Sector LLMs}

In this section, we propose recommendations
for how HCI researchers can engage with public sector practitioners to design LLMs that support their work.

\begin{itemize}
    \item \textbf{We encourage collaborations with practitioners to identify areas of public sector work that can benefit from the technical capabilities of LLMs and areas where these technologies should \textit{not} be used early on in the AI development process \cite{baumer11}.} Oftentimes, AI harms and failures are attributed to idealistic problem formulations, where there is insufficient consideration of how AI tools can create value for practitioners \cite{Saxena_mismatch2025, mothilal24}. Through speculative co-design workshops \cite{marji23}, applications of innovative risk matrices \cite{ Saxena_mismatch2025}, and exploratory research collaborations such as ours, practitioners and other stakeholders can identify contexts where resistance to automation is justified and where human discretion is necessary to augment transparent and accountable decision-making.

    \item \textbf{Through participatory methods, future research could identify and evaluate which forms of human–AI collaboration enhance workers' meaningfulness of public sector work while appropriately delegating tasks to LLM tools to improve overall performance. }Public sector and industry discourse on LLMs have largely emphasized its ability to improve the quality of public services delivered and efficiency gains but have paid less attention to workers' experiential aspect of working with AI tools \cite{soulofwork2024, cohere, brookings}. Sadeghian et al. \cite{soulofwork2024} recently found people prefer to work interactively with AI tools as it gives them a sense of meaningfulness in their work but that AI tools may be better at some tasks than humans. Our agency co-authors have suggested that LLMs could be used as an interactive deliberative tool \cite{Ma2025} in assisting workers document consistent and comprehensive details on cases (Section \ref{sec:rq3_pt1}). Further work can examine how to design tools that enhance the experiential experience of workers and clients.

    \item \textbf{Train agency leaders and frontline staff on evaluating LLM tool performance, and work toward building consensus across managerial levels on the tool’s evaluation criteria \cite{Saxena_mismatch2025, Kawakami_2024}.} For example, HCI researchers can co-develop customized toolkits with practitioners that allow workers to assess how LLM tools add value to their work. Instead of traditional methods of evaluating AI based on accuracy metrics, these toolkits should examine tradeoffs that can emerge between multiple evaluation metrics, such as model performance, data quality, and errors \cite{Saxena_mismatch2025}. Greater literacy on how to identify AI issues will allow workers to contest AI-driven decisions and reduce over-reliance on these tools \cite{karusala_contestability24}. Furthermore, our study's findings showed that management and frontline workers can differ in what they consider to be Activity-relevant. Accordingly, such evaluation toolkits should be co-developed through organization-wide involvement, including leadership teams and frontline staff.

\end{itemize}

\subsection{Limitations and Future Work}
Our study only draws on casenotes from one child welfare agency in Canada, so our findings may not be generalizable across other child welfare systems in North America, which are governed and regulated by other laws and policies. Even so, we believe our methodological approach can be applied across other child welfare systems, as most agencies draw up case plans that outline goals that a family will work towards and document regular casenotes. Furthermore, it is important to remember that Regular casenotes contain caseworker perceptions and may reflect worker biases and omit information pertinent to signaling case progress. Future work can examine how our study findings align and differ across other child welfare systems, as well as other welfare systems. Furthermore, in our work, we applied Service Plan Activities to track case progress in Regular casenotes. Following Ministry guidelines, these goals are standardized in language and are intended to address child welfare concerns surrounding family/child safety, permanency, and well-being. Further research should investigate how comprehensively these goals address clients’ child welfare concerns and if technical interventions can improve how goals are defined and documented by workers. Lastly, we only applied BERTopic and Llama 3.1 in our analysis. Although we experimented with other LocalLLM models, the final analysis was conducted using only Llama 3.1. Prior research has shown that LLMs and topic models may yield different outputs depending on the model \cite{paula25, el2024comparative}, and LLM performance may be affected by prompt brittleness issues \cite{promptbrittleness}. Future work should validate our findings by comparing the performance of additional models and experimenting with different prompt parameters. 

\section{Conclusion}

We conducted a study in collaboration with a large Canadian child welfare agency to examine how computational tools can be used to infer child welfare progress. Using two types of child welfare documents, a Service Plan that outlines case management goals for families and Regular casenotes that detail day-to-day information pertaining to a family’s case, we explored how AI tools can support caseworkers in tracking case progress. Our findings show that LLMs can assist caseworkers trace thematic trajectories in cases. However, we find LLMs cannot correctly identify which casenotes contain information relating to child welfare goals, especially as cases become more complex, because discretionary decision-making founded on social work practices is needed. Overall, we argue that LLMs should not be used to supplant caseworker decision-making but as decision-aid tools.

\begin{acks}
This research was supported by the NSERC Discovery Early Career Researcher Grant RGPIN-2022-04570, MITACS Accelerate program, and the University of Toronto Data Science Institute Summer Undergraduate Data Science Opportunities Program. Opinions, findings, and conclusions expressed in this paper are those of the authors. We sincerely thank our collaborators and anonymous reviewers whose suggestions and comments helped improve this manuscript.
\end{acks}

\bibliographystyle{ACM-Reference-Format}
\bibliography{sample-base}


\appendix

\section{Example topics grouped for thematic similarity}
\label{appendix:example_theme}
In Regular casenotes, we found that the topic model identified four topics related to legal matters. However, the topics differed in their mode of delivery, where one topic picked up on workers' descriptions of legal proceeding updates, while the other three topics picked up on legal updates written in the form of emails, texts, and phone calls. Similarly, in Service Plan Activities, we found four topics were related to substance abuse, wherein one focused on a person `not' being under the influence, while other topics focused on a person receiving specific treatments and services for alcohol or substance abuse. 

\section{Exemplar sentences for casenote themes}
\label{appendix:super_topic_desc}
This section provides exemplar sentences for the themes, shared and unique, across the different narrative types. All examples provided have been anonymized.

\begin{itemize}
  \item\textbf{(Shared Theme) Family Relationship/ Visits \& Parenting} 

\begin{myquote}
    \small{
    \textit{\textbf{Objective:} “Parent/caregiver has sufficient communication skills.”}\\
    \textit{\textbf{Activity:} “[name] and [name] will be connected to family counselling/ parenting support program”}\\
    \textit{\textbf{Regular Casenote:} “I informed dad of placement. l advised [name] by text message that [name] was placed in foster care today and that I will be in touch with him about access visits next  week. [name] responded saying thanks for the information and help.”}
    }
\end{myquote}

\item\textbf{(Shared Theme) Child Custody \& Criminal/Legal}

\begin{myquote}
    \small{
    \textit{\textbf{Objective:} “The child/youth does not engage in unlawful activity.”}\\
    \textit{\textbf{Activity:} “[name] and [name] to attend mediation to work out a custody arrangement for [name].”}\\
    \textit{\textbf{Regular Casenote:} “I attended the home, [name] was home with her husband. [name] advised that [name] is sleeping, it seems she is tired, she was at the day care for the all day… I served [name] with court documents. I asked her if she would like me to review the papers with her but she said that she is fine. We talked about court, [name] is ready for mediation and full custody to continue to care for [name]. ”}
    }
\end{myquote}

\item\textbf{(Shared Theme) Medical/Mental Health}

\begin{myquote}
    \small{
    \textit{\textbf{Objective:} “Child/youth receives preventative medical, dental and/or vision care.”}\\
    \textit{\textbf{Activity:} "[name] will take [name] for regular check up with the family doctor”}\\
    \textit{\textbf{Regular Casenote:} “Voicemail from walk in doctor - scratch may be old, it's not tender  - no bruising and no bruising elsewhere  - he is not in any distress  - pupils are good  - nothing in his ears  - no marks of abuse  - just wanted to follow up  - looks comfortable, alert and bright  - no worries at this time. I called Dr. [name] and he stated what he said in the voicemail. He said he had no worries or concerns and was not sure why he was seeing  [name].”}
    }
\end{myquote}

\item\textbf{(Shared Theme) Anger Management Conflict \& Safety}
\begin{myquote}
    \small{
    \textit{\textbf{Objective:} “Child/youth will not be exposed to physical conflict/violence in the home.”}\\
    \textit{\textbf{Activity:} “[name] and [name] will not expose [name] to partner/adult conflicts and issues are resolved amicably”}\\
    \textit{\textbf{Regular Casenote:} “[name] said she has always paid attention to her daughters, never let anyone carry her daughters when they were young, never left them alone  with anyone.”}
    }
\end{myquote}

\item\textbf{(Shared Theme) Administration related tasks, resources, and scheduling}

\begin{myquote}
    \small{
    \textit{\textbf{Objective:} “To connect [name] to appropriate community supports and services”}\\
    \textit{\textbf{Activity:} “The worker will make referrals to youth services to address [name]'s physical and mental health conditions”}\\
    \textit{\textbf{Regular Casenote:} “Call to victim witness assistance to obtain info. Spoke to a rep who advised [name]'s assigned worker is [name]. I advised her [name] only speaks [language] so won't be able to speak without an interpreter. She said she will update the file, send [name] an email to contact [name] with an interpreter. ”}
    }
\end{myquote}

\item\textbf{(Unique Theme) Attempts to Contact}

\begin{myquote}
\small{
\textit{\textbf{Regular Casenote:} “Unsuccessful scheduled phone call with [name]: she did not call. [name] and I were scheduled to have a phone call to get an update about how things are going with her and her access, and to discuss her plan and how that is progressing.”}
}
\end{myquote}

\item\textbf{(Unique Theme) Support Network}
\begin{myquote}
    \small{
    \textit{\textbf{Objective:} “For the family to have a large support network surrounding them that does not include the society.”}\\
    \textit{\textbf{Activity:} “For there to be a suitable network of supports to access as needed to help [name] and the family during this challenging time.”}
    }
\end{myquote}

\item\textbf{(Unique Theme) Child Development}

\begin{myquote}
    \small{
    \textit{\textbf{Objective:} “Child/youth's physical and cognitive skills are age appropriate.”}\\
\textit{\textbf{Objective:} “Child/youth demonstrates adequate social skills.”}\\
    }
\end{myquote}

\end{itemize}

\section{Analysis of Service Plan Objectives and Activities}
\label{appendix:obj_act_desc}

When we manually examined our 38-topic model solution for Service Plan Objectives, we found over half of the Objectives in our dataset (74\%, 1242/1677) were identically worded goals such as "Family resolves conflict regarding cultural differences," "Family engages with a strong support systems" or "Parent/caregiver physical or mental health does not affect parenting, family functioning and/or resources." The remaining Objectives included specific goals intended to be carried out by specific people or detailing specific actions, such as "[name] ensures [name]'s medical needs are met" or "To continue to try to locate biological mother," but even these objectives fell under consistent themes listed in Table \ref{tab:r1}. These findings suggest that high-level goals (i.e., Objectives) in child welfare at the agency are relatively standardized across the agency. On the other hand, manual inspections of our 82-topic model solution for Activities revealed how caseworkers provide customized supports to families based on individual circumstances so that a family may successfully meet their Service Plan's Objectives. There were more Activities in our dataset compared to Objectives because there are multiple paths for a family to achieve a Service Plan's Objectives. Moreover, Activities provided concrete, actionable steps to achieve a case's Objective. Each Activity included information on \textbf{1)} the person(s) the activity should be carried out by E.g., a child, parent, or caseworker. And \textbf{2)} required action to complete the activity. This could be a single, multiple/ongoing behavioral actions or an absence of an action, such as a child regularly attending school or a father not drinking alcohol in the presence of children. 

\section{Ambiguities in case progress}
\label{appendix:case_ambg}
The following bullet points provide a few examples of cases in which case progress toward Service Plan Activity completion was ambiguous. 

\begin{itemize}
\item When an activity is to, \textit{“[Name] to attend sobriety treatment programs”}, it was unclear if a caseworker attempting to refer and schedule the client to sobriety treatment programs would be considered Activity-relevant. 

\item In one case, a child was not enrolled in school. An Activity for the family included, \textit{“Conduct academic assessments for school reading, writing literacy for [name].”} The caseworker assisted the family in enrolling the child in the school so that they could take the academic assessments. In this case, it was unclear if school registration facilitation would be considered Activity-relevant. 

\item A family’s Activity stated that, \textit{“[name] to demonstrate parenting interactions and activities that match with the developmental age of her children.”} In this case, a bio-parent had access to visits with a child who was not living with the parent. Regular casenotes showed that the bio-parent would cancel scheduled access visits with the child, which occurs commonly across all child welfare cases when parents face unexpected scheduling conflicts. The first few cancellations appeared harmless, but it became increasingly clear that the bio-parent was cancelling many pre-scheduled access visits and not meeting the Service Plan’s Activity. In this case, our co-authors at the agency agreed that the cancellations can indicate Activity-relevant signals but one would only know this retrospectively after several cancellations had been made.

\end{itemize}

\end{document}